\documentclass[conference]{IEEEtran}
\IEEEoverridecommandlockouts
\usepackage{cite}
\usepackage{soul}
\usepackage{amsmath,amssymb,amsfonts}
\usepackage{algorithmic}
\usepackage{graphicx}
\usepackage{textcomp}
\usepackage{multirow}
\usepackage{url}
\usepackage[normalem]{ulem}
\usepackage{makecell}
\usepackage{array} 
\usepackage{multirow, hhline}

\usepackage[table]{xcolor} 
\def\BibTeX{{\rm B\kern-.05em{\sc i\kern-.025em b}\kern-.08em
    T\kern-.1667em\lower.7ex\hbox{E}\kern-.125emX}}
\begin{document}

\title{Efficient Prompt Tuning for Hierarchical Ingredient Recognition}


\vspace{-1.5em}
\author{%
  Yinxuan Gui \textsuperscript{1, 2}
  Bin Zhu \textsuperscript{2} 
  Jingjing Chen \textsuperscript{1,  \dag} \thanks{$^\dag$ Corrsponding author.} 
  Chong-Wah Ngo \textsuperscript{2} \\
 \\

 \textsuperscript{1} Shanghai Key Lab of Intell. Info. Processing, School of CS, Fudan University, China\\
 \textsuperscript{2} Singapore Management University \\
 yxgui22@m.fudan.edu.cn, \{binzhu, cwngo\}@smu.edu.sg, chenjingjing@fudan.edu.cn 
 }

\maketitle

\begin{abstract}
Fine-grained ingredient recognition presents a significant challenge due to the diverse appearances of ingredients, resulting from different cutting and cooking methods. While existing approaches have shown promising results, they still require extensive training costs and focus solely on fine-grained ingredient recognition. In this paper, we address these limitations by introducing an efficient prompt-tuning framework that adapts pretrained visual-language models (VLMs), such as CLIP, to the ingredient recognition task without requiring full model finetuning. Additionally, we introduce three-level ingredient hierarchies to enhance both training performance and evaluation robustness. Specifically, we propose a hierarchical ingredient recognition task, designed to evaluate model performance across different hierarchical levels (e.g., chicken chunks, chicken,  meat), capturing recognition capabilities from coarse- to fine-grained categories. Our method leverages hierarchical labels, training prompt-tuned models with both fine-grained and corresponding coarse-grained labels. Experimental results on the VireoFood172 dataset demonstrate the effectiveness of prompt-tuning with hierarchical labels, achieving superior performance. Moreover, the hierarchical ingredient recognition task provides valuable insights into the model's ability to generalize across different levels of ingredient granularity.
\end{abstract}

\begin{IEEEkeywords}
Hierarchical ingredient recognition, prompt tuning and vision-language model
\end{IEEEkeywords}

\begin{figure*}[htbp]
  \includegraphics[width=1\textwidth]{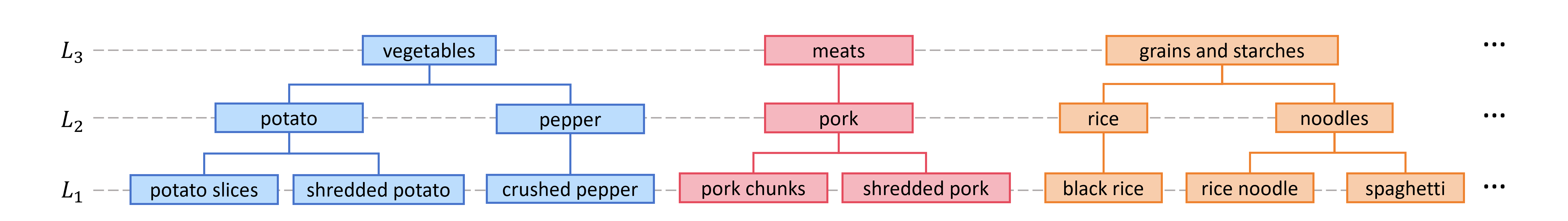}
  \vspace{-1.5em}
  \caption{The examples of our proposed three-level ingredient hierarchy.} 
  \label{fig:hierarchy}
  \vspace{-0.5em}
\end{figure*}

\section{Introduction}
\label{sec:intro}

With the growing pursuit of healthy diet and life, food computing~\cite{min2019survey-food-computing} is gaining increasing attention and numerous tasks have been explored, such as food classification~\cite{bossard2014food-classification-3, min2020isia-food-classification-5, yin2023foodlmm-food-classification-6}, ingredient recognition~\cite{chen2020study-ingr-recognition-1, chen2020zero-ingr-recognition-2, bolanos2017ingr-recognition-3, liu2020food-joint-learning}, recipe retrieval~\cite{zhu2019r2gan-retrieval-1, chen2017cross--retrieval-2, salvador2021revamping-retrieval-3, song2025enhancing-retrieval-4} and recipe generation~\cite{wang2022learning-recipe-generation-2, chhikara2024fire-recipe-generation-3, liu2024retrieval-generation-4}. In particular, ingredient recognition plays a critical and fundamental role in food-related research and applications~\cite{min2019ingredient-guide-research-1, yanai2014cooking-ingr-guide-research-2, gui2024navigating-ingr-guide-research-3}. 

Fine-grained ingredient recognition is a challenging task in nature. On the one hand, the same ingredient exhibits various visual appearances under different cooking and cutting methods, such as ``shredded pepper" and ``crushed pepper". On the other hand, different ingredients often have visual similarities, for example, ``hob blocks of carrot" and ``pumpkin chunks" have similar colors and shapes. To distinguish these ingredients, the existing works have explored ranging from region-wise features~\cite{chen2020study-ingr-recognition-1, luo2023ingredient-region-wise-1, gao2022dynamic-region-wise-2}, multi-task learning~\cite{liu2020food-joint-learning} to the relation among ingredients \cite{chen2020zero-ingr-recognition-2} for ingredient recognition. 
Nevertheless, these approaches only focus on fine-grained ingredients while ignoring the  hierarchical relationships among ingredients, for instance,
``shredded pepper" and ``crushed pepper" can be grouped into ``pepper", and ``hob blocks of carrot" and ``pumpkin chunks" belong to ``carrot" and ``pumpkin" respectively. Although~\cite{chen2020zero-ingr-recognition-2} considers ``is a” relationship among ingredients which shares similar spirit to the hierarchical relationship in this paper, for example, “bell pepper” is a “chili”.  However, the relationship is constrained to a specific range of ingredients, without introducing new coarse-grained ones to establish a fine-to-coarse hierarchy. 
Additionally, most existing methods rely on fully fine-tuning models, which incurs significant training costs. Prompt tuning is an efficient way to adapt pretrained models to downstream tasks. Although it has been extensively explored~\cite{zhou2022learning-CoOp, zhou2022conditional-CoCoOp, lu2022prompt-tuning-distribution, huang2022unsupervised-prompt-tuning}, it is primarily focused on single-label scenarios rather than multi-label ones~\cite{sun2022dualcoop, guo2023texts-as-images-prompt-tuning} and has not been explored for ingredient recognition yet.
In addition, the pretrained vision-language models~\cite{radford2021learning-CLIP} and large multimodal models (LMMs)~\cite{zhu2023minigpt-LMM1, liu2024visual-llavaLMM3}  have demonstrated excellent zero-shot capabilities powered by the massive training data. However, these models tend to produce coarse prediction for ingredient recognition and often struggle to deal with very fine-grained ingredient recognition. 

To address these limitations, we propose a hierarchical ingredient recognition which is designed to evaluate model performance across different hierarchical levels (e.g. crushed pepper, pepper, vegetables). We first construct three-level ingredient hierarchies from fine-grained to coarse-grained based on VireoFood172~\cite{chen2016deep-vireofood172}, a dataset including fine-grained ingredient labels. In addition, we introduce the hierarchies to enhance both training performance and evaluation robustness. Specifically, for the former, we propose a two-stage cross-hierarchy training method based on efficient prompt tuning, which adapts
pretrained visual-language models (VLMs) (e.g., CLIP~\cite{radford2021learning-CLIP}) to
the ingredient recognition task without requiring full model
fine-tuning. Our method first trains models for ingredient recognition at different levels separately in the first stage, and then we leverage implicit hierarchical relationships to train models with both fine-grained and coarse-grained ingredient labels in the second stage.
We evaluate our method on VireoFood172~\cite{chen2016deep-vireofood172} dataset and the experiment results demonstrate the effectiveness of leveraging ingredients of different hierarchies. We further evaluate two pretrained VLMs, CLIP~\cite{radford2021learning-CLIP} and LLaVA~\cite{liu2024visual-llavaLMM3}, for zero-shot hierarchical ingredient recognition, offering insights into their ability to generalize across varying levels of ingredient granularity.




\section{Related Work}
\subsection{Prompt tuning for visual-language models}
Visual-language models, such as CLIP~\cite{radford2021learning-CLIP}, have shown remarkable capabilities on several tasks. To transfer knowledge from pretrained models to downstream tasks, prompt tuning has become a popular method without requiring full model finetuning.  In \cite{zhou2022learning-CoOp}, CoOp is proposed as a prompt learning-based approach and outperforms manually designed prompts. Based on it, \cite{zhou2022conditional-CoCoOp}  further proposes CoCoOp to learn generalizable prompts by generating for each image an input-conditional token. \cite{lu2022prompt-tuning-distribution} proposes a novel prompt learning method to handle the varying visual representations by learning distribution of diverse prompts. Huang et al.~\cite{huang2022unsupervised-prompt-tuning} presents an unsupervised prompt learning approach to deal with a scenario, in which the labels of datasets are unavailable. Different from the aforementioned works, DualCoOp~\cite{sun2022dualcoop} firstly adapts CLIP to multi-label recognition task by learning a pair of negative and positive prompts to achieve binary classification for each class. ~\cite{guo2023texts-as-images-prompt-tuning}  aligns the images modality and text modality to treat text descriptions as images for prompt tuning because texts are easier to collect. We introduce the prompt tuning method into ingredient recognition, enabling the utilization of CLIP's knowledge without extensive training costs. To adapt to the multi-label ingredient recognition task, we employ DualCoop~\cite{sun2022dualcoop} as our prompt tuning approach.

\subsection{Ingredient recognition}
Ingredient recognition is a challenging task because ingredients exhibit various appearances and are small in size.  Existing works on ingredient recognition mainly focus on fully fine-tuning CNNs~\cite{szegedy2015going-vgg, he2016deep-resnet, szegedy2016rethinking-inception}. For example, \cite{bolanos2017ingr-recognition-3} empolys ResNet-50~\cite{he2016deep-resnet} and InceptionV3~\cite{szegedy2016rethinking-inception} which are
 pretrained on ILSVRC2012. Since ingredients are closely related to other forms of food data, such as food categories and recipes, some works improve ingredient recognition performance in multi-task~\cite{chen2016deep-vireofood172, liu2020food-joint-learning, chen2020study-ingr-recognition-1} manner.
Recently, 
~\cite{gao2022dynamic-gaojx-ingr} proposes D-Mixup to boost the recognition performance by formulating the task as a long-tailed classification problem based VGG and ResNet~\cite{he2016deep-resnet} backbone. In \cite{luo2023ingredient-region-wise-1}, a model is trained on ResNet to mitigate the negative impact of complex image background and imbalanced ingredient classes. Food large multi-modal models~\cite{yin2023foodlmm-food-classification-6, jiao2024rode} have been explored recently for multiple tasks in food domain, including ingredient recognition and beyond, showing promising results. However, these approaches require extensive training costs due to full fine-tuning. Few works utilize prompt tuning methods to solve ingredient recognition task, especially as it is a multi-label task, which is less commonly addressed by prompt tuning.  Additionally, we not only utilize the fine-grained ingredients in the dataset, but also the relationship among ingredients with different granularity by constructing an ingredient hierarchy. We introduce the hierarchies to train models with both coarse- and fine-grained ingredients and evaluate models with higher robustness.

\begin{figure*}[htbp]
  \includegraphics[width=1\textwidth]{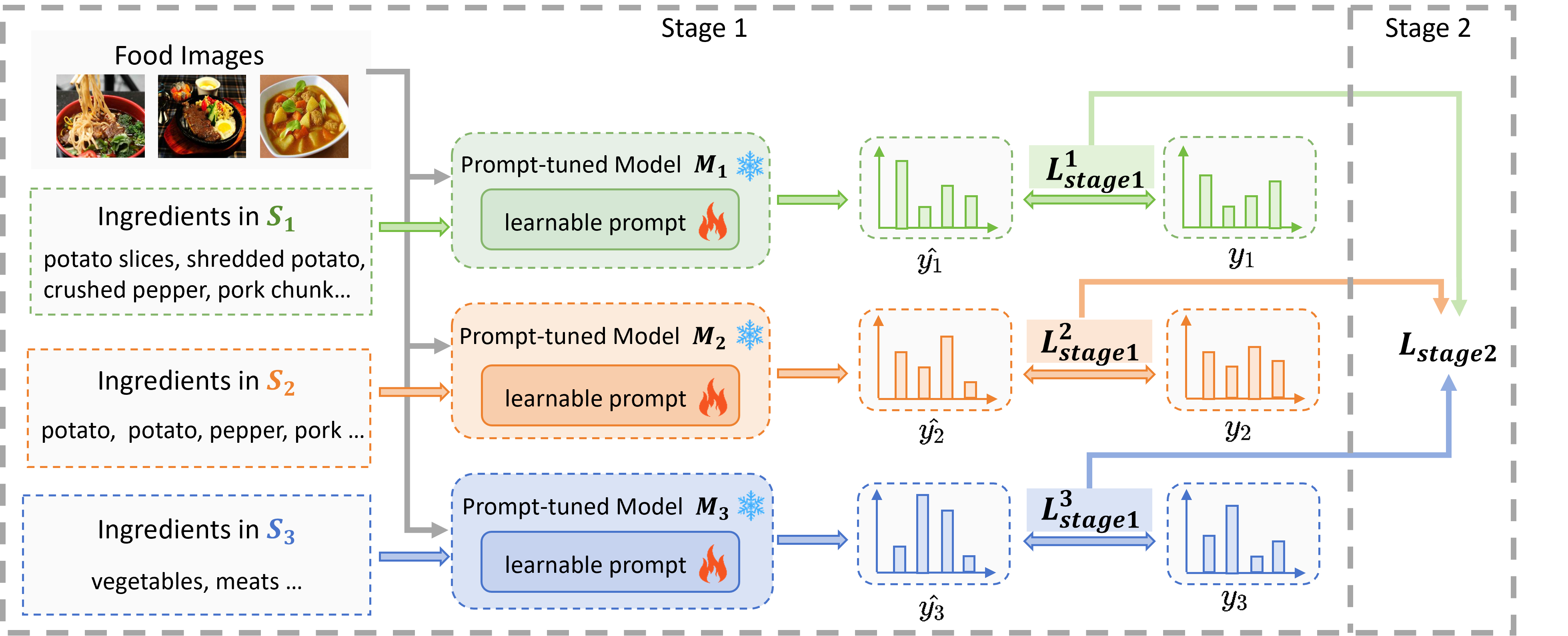}
  \caption{The overview of our proposed two-stage cross-hierarchy training method. In the first stage, we train three prompt tuning models of different hierarchies separately. In the second stage, the losses are combined as $L_{stage2}$ to train three models together, leveraging the ingredient hierarchy.}

  \label{fig:model-overview}
  \vspace{-0.2cm}
\end{figure*}

\section{Method}

\subsection{Ingredient hierarchy construction}
We construct a three-level ingredient hierarchy $H$, where the $i$-th level denoted as $L_i$, each level is composed of an ingredient labels set $S_i$, $i \in \{1, 2, 3\}$. 
From levels 1 to 3, the set is constructed from
fine-grained to coarse-grained ingredients.  
Fig.~\ref{fig:hierarchy} presents the structure of $H$, with some examples of ingredients at different levels. For example, fine-grained ingredients prepared with specific cooking methods in $S_1$, such as ``potato slices" and ``shredded pork", are grouped into broader ingredients like ``potato" and ``pork" in $S_2$, which are then further grouped into more coarse-grained ingredients such as vegetables and meats in $S_3$. Most ingredients have a three-level hierarchy, but for a small portion consisting of two levels, we repeat the ingredient of $L_1$ at $L_2$.
We first employ Claude 3.5 Haiku~\cite{claude}, a remarkable large language model, to generate the hierarchy and then carefully conduct manual refinement. Specifically, we complement some missing ingredients by large language model and group some ingredients together, for example, we group ``pickled red peppers" to ``pepper" which is grouped into other categories originally.

Given a food image $x$ that contains a set of ingredients $y$, we can obtain hierarchical ingredient labels $y_1$, $y_2$ and $y_3$ with different granularity based on $H$, where each ingredient in $y_i$ belongs to $S_i$. For example, given ``crushed pepper" as a label, the labels are denoted as ``crushed pepper", ``pepper" and ``vegetables" at three levels.
We introduce the hierarchical ingredient labels to improve training performance of pretrained visual language models, which is presented in Section~\ref{sec:training}.

\subsection{Cross-hierarchy prompt tuning}
\label{sec:training}
 We propose an efficient two-stage prompt tuning method based on pretrained visual language model (e.g. CLIP~\cite{radford2021learning-CLIP}) without training the whole model. Fig.~\ref{fig:model-overview} depicts an overview of our method. We follow DualCoOp~\cite{sun2022dualcoop} in the first stage. Given food images and hierarchical ingredient labels, three CLIP models, $M_1$, $M_2$ and $M_3$, are prompt-tuned using $S_1$, $S_2$, and $S_3$ as labels respectively. The process of the first-stage training can be formalized as follows:


\vspace{-1.0em}
\begin{align}
\label{eq:stage1-training loss}
\begin{split}
  &\hat{y_i} = M_i(x, S_i), \\
  & L_{stage1}^{i} = \mathcal{L}(y_i, \hat{y_i}), i \in \{1,2,3\},
\end{split}
\end{align}

\noindent where $x$ is a food image, $L_{stage1}^{i} $ is the loss value of model $M_i$ and the $\mathcal{L}$ is the loss function. In particular, we employ DualCoOp~\cite{sun2022dualcoop} as our prompt tuning method and adopt the Asymmetric Loss~\cite{ridnik2021asymmetric} as $\mathcal{L}$.

However, the first training phase fails to represent the hierarchical connections between labels at various levels.
To model the connections, we further introduce the hierarchy to train the three models together in the second stage, supervised by a combination of their individual losses:
\begin{equation}
\label{eq:stage2-training loss}
\begin{split}
  L_{stage2} = \lambda_{1}L_{stage1}^{1}  + \lambda_{2}L_{stage1}^{2}  + \lambda_{3}L_{stage1}^{3} , \\
\end{split}
\end{equation}
where $\lambda_{1}$, $\lambda_{2}$, and $\lambda_{3}$ are hyper-parameters to balance losses of different hierarchies.

Note that the three models are frozen except for the prompt parameters, which are trainable in both the first and second stages. This design significantly improves efficiency by avoiding training whole models and reducing computational costs.

\begin{table}
\caption{Statistics of the number of ingredients at each hierarchy level.}
\vspace{-0.5em}
\label{tab:number-of-ingredients}
\begin{center}
\begin{tabular}{l|  >{\columncolor{red!5}}c  >{\columncolor{blue!5}}c  >{\columncolor{yellow!5}}c}
    \hline
    Level of the hierarchy&1&2&3\\
    \hline
    Number of ingredients&353&138&13\\
    \hline

   \hline

\end{tabular}
\end{center}
\vspace{-2em}
\end{table} 



\begin{table*}
\renewcommand{\arraystretch}{1.2} 
\caption{Performance comparison with existing methods on VireoFood172 dataset. The fully finetuning methods are marked in gray for reference. \#P represents the number of trainable parameters.}
\vspace{-2em}
\label{tab:main-performance}
\begin{center}
\resizebox{\linewidth}{!}{
\begin{tabular}{c | c|>{\columncolor{red!5}}c >{\columncolor{red!5}}c >{\columncolor{red!5}}c >{\columncolor{red!5}}c >{\columncolor{red!5}}c|>{\columncolor{blue!3}} c >{\columncolor{blue!3}}c >{\columncolor{blue!3}}c >{\columncolor{blue!3}}c >{\columncolor{blue!3}}c|>{\columncolor{yellow!5}} c >{\columncolor{yellow!5}} c >{\columncolor{yellow!5}}c >{\columncolor{yellow!5}}c >{\columncolor{yellow!5}}c}
    \hline
    \multicolumn{2}{c|}{Level}&\multicolumn{5}{c|}{ \cellcolor{red!5}1}&\multicolumn{5}{c|}{\cellcolor{blue!3}2}&\multicolumn{5}{c}{\cellcolor{yellow!5}3}\\
    \hline
    \multicolumn{2}{c|}{Metric} & P & R & IOU & F1 &  \#P & P & R & IOU & F1 &  \#P & P & R & IOU  & F1  &  \#P\\
    \hline

    \multicolumn{2}{c|}{CLIP~\cite{radford2021learning-CLIP} zero-shot} & 5.85 & 11.61  & 5.11& 7.78 & - & 16.74 & 23.38 & 13.47 & 19.51 & - & 44.30 & 51.06   & 35.61 & 47.44 & - \\
    \hline

    \multicolumn{2}{c|}{DualCoOp~\cite{sun2022dualcoop}  (first stage)} & 61.60
    & 69.02 & 54.42  & 65.10 & \textbf{5.8M} & 64.01 & 73.69  & 57.59  & 68.51 & \textbf{2.3M} & 72.62 & 85.45  & 69.73 & 78.51 & \textbf{0.2M}\\
    \hline
    \multicolumn{2}{c|}{\textbf{Ours} (second stage)}  & 
    \textbf{62.88} & \textbf{69.17}  & \textbf{55.39} & \textbf{65.88}  & \textbf{5.8M} &\textbf{64.43} &73.54 &\textbf{57.80}  & \textbf{68.68}& \textbf{2.3M} & \textbf{72.85} & 85.35  & \textbf{69.86} & \textbf{78.60}& \textbf{0.2M} \\
   \hline
   %
    \rowcolor{gray!10}
   & Arch-D~\cite{chen2016deep-vireofood172} & - & - & -   & 67.17 & 130M &\multicolumn{5}{c|}{-} & \multicolumn{5}{c}{-}\\
   \cline{2-17}

   \rowcolor{gray!10} 
    & AFN+BFL~\cite{liu2020food-joint-learning} & - & -  & - & 73.63 & 61M &\multicolumn{5}{c|}{-} & \multicolumn{5}{c}{-}\\
    \cline{2-17}

   \rowcolor{gray!10} 
   \multirow{-3}{*}{\makecell{fully \\finetuning\\ methods}} &  CACLNet~\cite{luo2023ingredient-region-wise-1} & - & -  & - & 79.28 & 45M &\multicolumn{5}{c|}{-} & \multicolumn{5}{c}{-}\\
     \hline
\end{tabular}}
\end{center}
\vspace{-1em}
\end{table*}


\begin{figure*}[htbp]
  \includegraphics[width=1\textwidth]{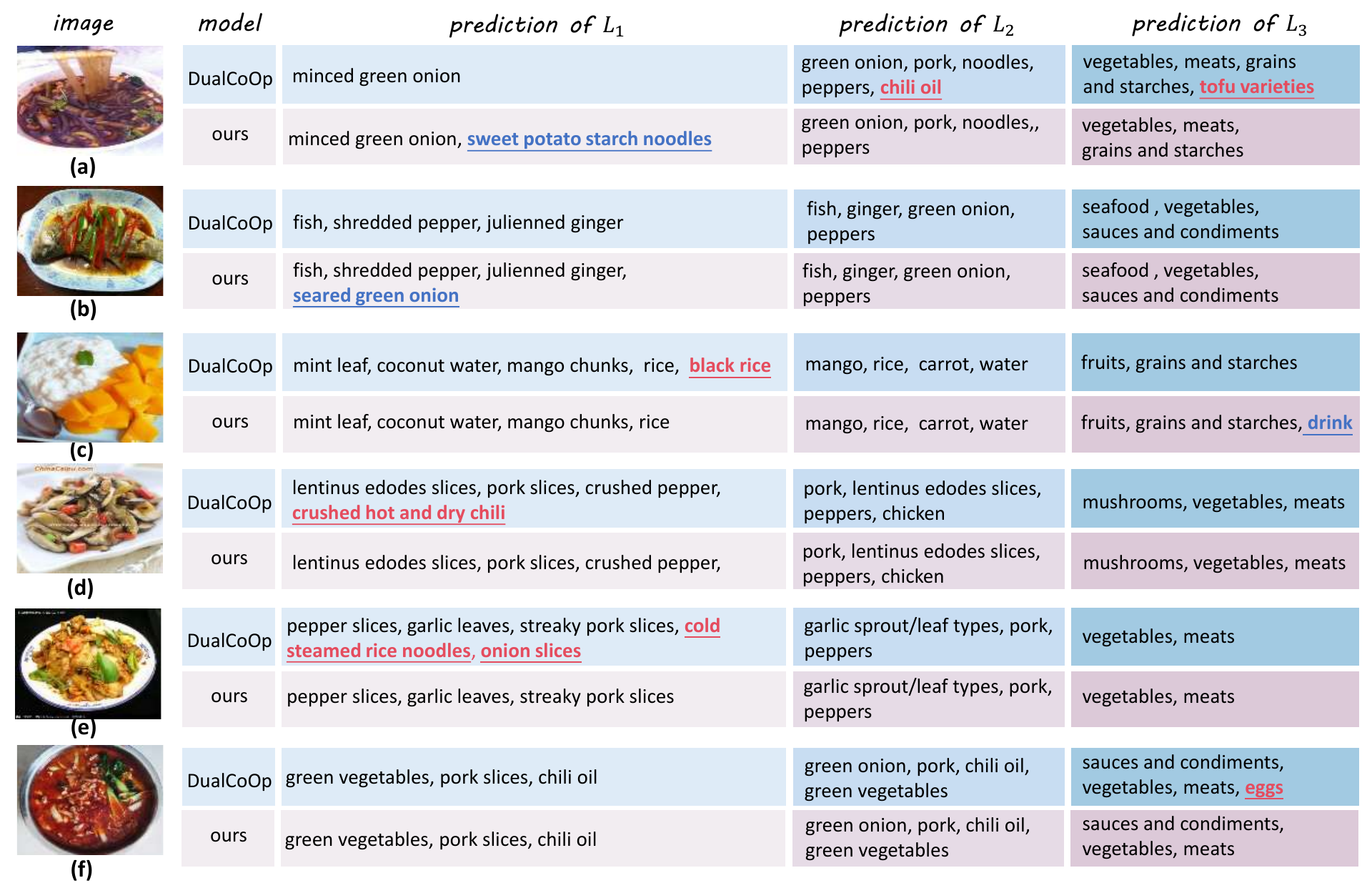}
  \caption{Qualitative examples of hierarchical ingredient recognition on VireoFood172. False negatives removed by our method are marked in red with underlines. Additionally, true positives complemented by our method are marked in blue with underlines.}
  \vspace{-0.5cm}
  \label{fig:examples-prediction-results}
  
\end{figure*}

\section{Experiments}
\subsection{Experiments setting}
\noindent \textbf{Dataset.}
We conduct experiments on VireoFood172\cite{chen2016deep-vireofood172} dataset, which consists of 353 fine-grained ingredient labels. 
We adopt 
the original data splits for training, validating and testing. A three-level hierarchy is constructed on these 353 ingredients and the number of ingredients at each level is shown in Table \ref{tab:number-of-ingredients}. The first level is the original ingredient label in the dataset. As far as we know, there is no other large-scale food datasets with image-level human annotated ingredients. 

\noindent \textbf{Two stage training.} Given the promising performance of DualCoOp\cite{sun2022dualcoop}, which adapts CLIP\cite{radford2021learning-CLIP} to multi-label image recognition, we choose DualCoOp as prompt-tuning baseline mentioned in Section~\ref{sec:training} and adopt its training strategy and loss function. In the first stage, we train three models at all levels separately for 110 epochs with the learning rate initialized as 0.002. In the second stage, three models are trained together for 60 epochs and the learning rate is initialized as 0.001.  The learning rate is always decayed by the cosine annealing rule. The $\lambda_1$, $\lambda_2$, and $\lambda_3$ in equation \ref{eq:stage2-training loss} are set as different combinations to explore the influence of the portions of hierarchy levels, which will be shown in Section\ref{ablation study}.

\noindent \textbf{Zero-shot evaluation.} We evaluate hierarchical zero-shot capability on two remarkable pretrained visual language models, LLaVA-1.5-7b~\cite{liu2024visual-llavaLMM3}
 and CLIP~\cite{radford2021learning-CLIP}. The ViT-B/32 is employed in CLIP's backbone.

\noindent \textbf{Evaluation metrics.}
As hierarchical ingredient recognition is a multi-label task, we employ precision (P), recall (R), intersection over union (IOU) and F1 score as our metrics. We also report the trainable parameters (\#P) of different methods to measure the computational complexity.

Table~\ref{tab:main-performance} presents the performance of our proposed method (second stage), our baseline DualCoOp~\cite{sun2022dualcoop} (first stage), along with zero-shot performance of CLIP and some fully fine-tuning methods. Due to the fine-grained ingredients such as ``crushed pepper" and ``shredded pepper", CLIP-zero-shot shows poor capability to distinguish them. With prompt tuning by DualCoOp in the first stage, the performance is significantly boosted than CLIP-zero-shot in all the three levels, with only 5.8M, 2.3M and 0.2M trainable parameters for levels 1, 2 and 3 respectively. Importantly, our method after the second stage training achieves superior performance compared to DualCoOp without increasing the trainable parameters, 
which demonstrates the relationships among different hierarchy levels can enhance the accuracy of hierarchical ingredient recognition consistently.
Though fully fine-tuning methods (highlighted in gray) show better results, the training cost is at least 7.7 times more than our method. 
Our method with efficient prompt tuning shows decent performance with much lower trainable parameters.

\begin{figure*}[htbp]
  \includegraphics[width=0.98\textwidth]{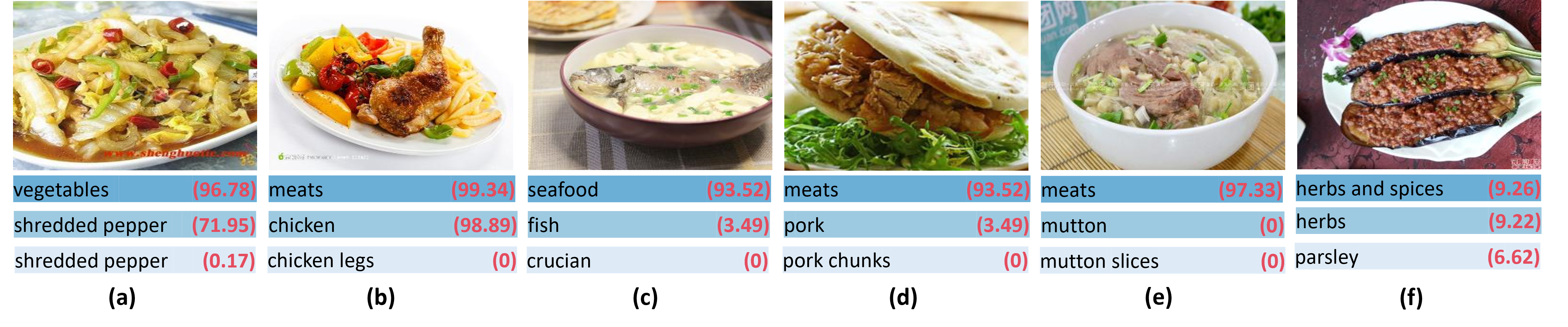}
   \vspace{-0.2cm}
  \caption{F1 score (\%) of hierarchical ingredient recognition at different levels based on zero-shot evaluation of LLaVA.}
  \label{fig:zero-shot-hier}
  \vspace{-0.3cm}
\end{figure*}

\vspace{-0.3em}
\subsection{Performance comparison}
\vspace{-0.25em}
Fig.~\ref{fig:examples-prediction-results} further shows a few qualitative examples which compare the prediction results of our method and DualCoOp across three levels of the hierarchy. On the one hand, our cross-hierarchy training method can complement true positives. 
For example, in Fig.~\ref{fig:examples-prediction-results} (a), ``sweet potato starch noodles" is not predicted at $L_1$ but predicted as ``grains and starches" and ``noodles" at  $L_3$ and $L_2$ by DualCoOp. Our method complements this by leveraging the implicit connection among ``grains and starches", ``noodles" and the ingredients belong to them at $L_1$. The same pattern can be observed as ``seared green onion" in Fig.~\ref{fig:examples-prediction-results} (b) and ``drink" in Fig.~\ref{fig:examples-prediction-results} (c), which are also predicted in other levels. On the other hand, our method can also remove false negatives, such as ``chili oil" and ``tofu varieties" in Fig.~\ref{fig:examples-prediction-results} (a), ``cold steamed rice noodles" and ``onion slices" in Fig.~\ref{fig:examples-prediction-results} (e) and ``eggs" in  Fig.~\ref{fig:examples-prediction-results} (f). 

\begin{table}
\renewcommand{\arraystretch}{1.4} 
\caption{Ablation study of hyperparameters of weights in different hierarchies in terms of IoU. $\Delta$ Avg $\uparrow$ indicates average improvement across the three levels.}
\vspace{-1.5em}
\label{tab:ablation}
\begin{center}
\resizebox{\linewidth}{!}{
\begin{tabular}{c | l |>{\columncolor{red!5}} c >{\columncolor{blue!3}}c >{\columncolor{yellow!5}}c c}
    \hline
    \multicolumn{2}{c|}{$L$}& 1 & 2 & 3 & $\Delta$ avg $\uparrow$\\
    \hline
    \multicolumn{2}{c|}{$stage_1$}& 54.42 & 57.59 & 69.73 & -\\
    \hline
    \multirow{3}{*}{\makecell{$stage_2$\\ $\{ \lambda_1, \lambda_2, \lambda_3\}$\\$\lambda_1+\lambda_2+\lambda_3=1$}} & $\{0.6, 0.25, 0.15\}$ & 55.39 & \textbf{57.80} & 69.86  & \textbf{0.47}\\
    \hhline{|~|-----|}
    & $\{0.7, 0.20, 0.10\}$ & 55.41 & 57.66  & 69.85  & 0.39\\
    \hhline{|~|-----|}
    & $\{0.8, 0.15, 0.05\}$ & 55.53 & 57.61 & \textbf{69.92} & 0.44\\
   \hhline{|~|-----|}
    & $\{0.9, 0.05, 0.05\}$ & \textbf{55.57} & 57.57 & \textbf{69.92} & 0.44 \\

   \hline
\end{tabular}}
\end{center}
\vspace{-1.5em}
\end{table}

An interesting observation is that our method can remove a false negative while retaining a true positive when both are similar ingredients that can be grouped into the same class. This finding is worthy because one of the challenges of fine-grained ingredient recognition is the difficulty in distinguishing similar but different ingredients. Considering $L_1$ which is most fine-grained, in Fig.~\ref{fig:examples-prediction-results} (c), ``black rice" and ``rice" are predicted simultaneously by DualCoOp, but after the training of our method, the former is retained as a correct prediction, while the latter is not predicted. Similarly, ``crushed pepper" is retained and ``crushed hot and dry chili" is removed in Fig.~\ref{fig:examples-prediction-results} (d). 


\subsection{Ablation study}
\label{ablation study}
In this section, we investigate the impact of hyperparameters $\lambda_1$,  $\lambda_2$, and $\lambda_3$ in Equation~\ref{eq:stage2-training loss}. We conduct experiments on four different combinations and the IOU performance is presented in Table~\ref{tab:ablation}. Generally, the training of the second stage always achieves better performance compared with the first stage, regardless of the portions of the three hierarchies. Specifically, the performance of each level is sensitive to value of corresponding $\lambda$, as the hyperparametes are designed to control significance of different levels. For example, when $\lambda_1$ is set as 0.9, which is the highest among four settings, the performance of $L_1$  shows the most significant improvement. Overall, when \{$\lambda_1$,  $\lambda_2$, $\lambda_3$ \} are set to \{0.6, 0.25, 0.15\}, the performance reaches its highest improvement. 



\begin{table}
\renewcommand{\arraystretch}{1.2} 
\caption{Zero-shot performance across different hierarchies.}
\vspace{-1.5em}
\label{tab:zero-shot}
\begin{center}
\resizebox{\linewidth}{!}{
\begin{tabular}{c | c |>{\columncolor{red!5}} c >{\columncolor{blue!5}}c >{\columncolor{yellow!5}} c |>{\columncolor{red!5}} c >{\columncolor{blue!5}}c  >{\columncolor{yellow!5}} c}
    \hline
    \multicolumn{2}{c|}{model}& \multicolumn{3}{c|}{LLaVA~\cite{liu2024visual-llavaLMM3}} & \multicolumn{3}{c}{CLIP~\cite{radford2021learning-CLIP}}\\
    \hline
    \multicolumn{2}{c|}{$L$}& 1 & 2 & 3 & 1 & 2 & 3\\
    \hline
    \multirow{4}{*}{metric} & P & 12.24 & 34.35 & 72.83 & 5.85 & 16.74 & 44.30 \\
    \hhline{|~|-------|}
    & R & 10.25 & 29.10 & 57.67 & 11.61 & 23.38 & 51.06 \\
    \hhline{|~|-------|}
    & F1 & 11.16 & 31.51 & 64.37 & 7.78 & 19.51 & 47.44 \\
    \hhline{|~|-------|}
    & IOU & 7.86 & 22.42 & 54.54 & 5.11 & 13.47 & 35.61 \\

   \hline
\end{tabular}}
\end{center}
\end{table}

\subsection{Zero-shot hierarchical ingredient recognition}
\label{evaluation}
Except for enhancing training performance, the ingredient hierarchy can also be utilized for model evaluation. Given zero-shot prediction results $P_1$ in which ingredients all belong to $S_1$, we can map them to the other two levels and get the corresponding prediction $P_2$ and $P_3$. We evaluate predictions at each level and table~\ref{tab:zero-shot} presents zero-shot results in VireoFood172~\cite{chen2016deep-vireofood172} based on LLaVA~\cite{liu2024visual-llavaLMM3} and CLIP~\cite{radford2021learning-CLIP}.
Both models exhibit relatively poor performance on the fine-grained labels in $L_1$. When the ingredients group together from fine- to coarse-grained, the performance improves significantly. To investigate the model's performance for specific ingredients at different levels, we compute F1 score of each ingredient at different levels based on LLaVA and show some examples in Fig.~\ref{fig:zero-shot-hier}. Typically, the models can not recognize fine-grained ingredients such as ``shredded pepper", ``chicken legs" and ``crucian" in Fig.\ref{fig:zero-shot-hier} (a), (b) and (c).
Additionally, when these ingredients are grouped into broader categories like ``pepper", ``chicken"  and ``fish", the performance achieves a significant improvement. For meats, the models usually fail to differentiate specific types of meat, categorizing them as meat simply instead, such as ``pork" and ``mutton" in Fig.\ref{fig:zero-shot-hier} (d), (e). However, for some very small ingredients, the model fails to make predictions at any level, such as ``parsley" in Fig.\ref{fig:zero-shot-hier} (f). These results illustrate the sensitivity of visual language models to ingredients at different levels of granularity, as well as the limitations of their recognition capabilities.

\section{Conclusion}
We have presented a novel hierarchical framework for ingredient recognition. By constructing a three-level ingredient hierarchy and employing a two-stage cross-hierarchy training strategy with efficient prompt tuning, we adapt pretrained VLMs to this task without full fine-tuning, achieving improved performance with reduced training costs. Experiments on the VireoFood172 dataset demonstrate the effectiveness of our approach in enhancing recognition across different granularities. Additionally, we evaluate the zero-shot capabilities of VLMs like CLIP and LLaVA, providing insights into their generalization potential for ingredient recognition. Our work advances ingredient recognition by integrating hierarchical information, offering a scalable, efficient method and highlighting opportunities for future research in food computing.

\section*{Acknowledgment}
This research/project is supported by the Ministry of Education, Singapore, under Academic Research Fund (AcRF) Tier 1 grant (No. MSS23C018) and Tier 2 (Proposal ID: T2EP20222-0046). Any opinions, findings and conclusions or recommendations expressed in this material are those of the authors and do not reflect the views of the Ministry of Education, Singapore.

\bibliographystyle{IEEEbib}
\bibliography{icme2025references}


\end{document}